\documentstyle[12pt,slhp3m]{article}
\begin{document}


\def\A{{\cal A}}
\def\B{{\cal B}}
\def\dim#1{\mbox{\,#1}}

\def\reffigSfour{1}
\def\reffigStwo{3}
\def\reffigGD{11}

\def\figdir{.}
\def\placefig#1{#1}

\title{Building a Cosmological Hydrodynamic Code: Consistency Condition,
Moving Mesh Gravity and SLH-P$^3$M} 
\author{Nickolay Y.\ Gnedin\altaffilmark{1,2} and 
Edmund Bertschinger\altaffilmark{1}}
\altaffiltext{1}{Department of Physics, Massachusetts Institute of Technology,
Cambridge, MA 02139; e-mail: \it gnedin@arcturus.mit.edu, 
http://arcturus.mit.edu/}
\altaffiltext{2}{Princeton University Observatory, Peyton Hall, 
Princeton, NJ 08544; e-mail: \it gnedin@astro.princeton.edu}


\begin{abstract}

Building a self-gravitating hydrodynamic code as a combination of a
hydrodynamic solver and a gravity solver is discussed. We show
that straightforward combining those two solvers generally leads to
a code that does not conserve energy locally, and instead a special
Consistency Condition ought to be satisfied.
A particular example
of combining Softened Lagrangian Hydrodynamics (SLH) with a
P$^3$M gravity solver is used to demonstrate the effect of
the Consistency Condition for a self-gravitating
hydrodynamic code. The need to supplement the SLH method with the
P$^3$M gravity solver arose because the Moving Mesh Gravity solver,
used in conjunction with the SLH method previously, was found to produce
inaccurated results. We also show that most existing
cosmological hydrodynamic codes implicitly satisfy the Consistency
Condition.

\end{abstract}


\def\capSF{
Dark matter particles in a subvolume of four 32$^3$ simulations: 
the original SLH $\Omega_b=0$ (dark matter only) simulation 
({\it lower left panel}),
the pure P$^3$M simulation
({\it lower right panel}), the original $\Omega_X=0$ (baryons only) simulation
({\it upper left panel}),
and the $\Omega_b=0$ SLH simulations with dark matter density smoothed
in quasi-Lagrangian space ({\it upper right panel}).}

\def\capSP{
Density profiles for four clusters shown in Fig.\reffigSfour: P$^3$M only
({\it solid line}), $\Omega_b=0$ SLH ({\it dotted line}), $\Omega_X=0$ SLH 
({\it short-dashed line}), and $\Omega_b=0$ SLH with smoothing 
({\it long-dashed line}). The radius has units of the mean interparticle 
spacing; the force softening is $R=0.1$.}

\def\capST{
Dark matter particles in a subvolume of two 64$^3$ simulations: the 
$\Omega_b=0$ SLH simulation with smoothing ({\it left panel}) and the 
pure $N$-body simulation ({\it right panel}).}

\def\capSQ{
Density profiles for two simulations shown in Fig.\reffigStwo: P$^3$M only
({\it solid line}) and $\Omega_b=0$ SLH with smoothing ({\it dotted line}).}

\def\capGF{
{\it Left panel}, Green functions for ten randomly chosen points in 
the 64$^3$ SLH simulation ({\it solid lines}) 
and the inverse square law ({\it thin 
dotted line}). {\it Right panel}, Green functions for two points chosen near
the center of the cluster from the left panel of Fig.\reffigStwo\ 
({\it solid
lines}) and the Green function for one of those points calculated by solving
a Poisson equation for three components of the gravitational force on the 
deformed mesh ({\it dashed line}); shown with the thin dotted line is the
inverse square law.}

\def\capPT{
Baryonic particles (cell centers) for two 32$^3$ SLH-P$^3$M simulations:
without Consistency Condition ({\it left panel}) and with Consistency
Condition ({\it right panel}).}

\def\capPQ{
Density profiles for two largest clusters in the 64$^3$ SLH-P$^3$M
$\Omega_X=0$ simulation for dark matter ({\it dashed line}) and baryonic
gas ({\it dotted line}); shown with the solid line is the density profile
for the same cluster from the pure $N$-body (P$^3$M only) 
simulation with the same
initial conditions.}

\def\capEE{
Total energy error for the SLH-P$^3$M\,64 run as a function of
scale factor $a$; the total error fluctuates around zero with amplitude
1\% (except for $a<0.2$ where the energy error is not computed accurately
enough).}

\def\capCC{
Comparison of integrated quantities: volume averaged 
temperature $\langle T\rangle$ ({\it upper left panel}), 
mass averaged temperature
$\langle T\rangle_\rho$ ({\it upper right panel}), 
X-ray luminosity $L_x$ ({\it lower left panel}, in
arbitrary units), and rms density fluctuations $\sigma$ 
({\it lower right panel}),
for the SLH-P$^3$M approach (solid circles connected with the bold solid line)
to the other existing codes: original SLH (stars connected with the bold
solid line), COJ ({\it filled squares}), TVD ({\it small filled circles}),
PSPH ({\it open circles}) and TSPH ({\it open triangles}).}

\def\capGS{
A slice of original (256$^3$ rebinned) data for SLH-P$^3$M\,64 ({\it a\,}),
and SLH-P$^3$M\,128({\it b\,}) runs. 
The top row shows density and the bottom row shows
temperature; the right column shows a zoom of a $1/16$ of the whole slice
($24<x<40$, $40<y<56$) containing one of the largest clusters present
in the slice.}

\def\capGD{
Zooms of the density contours of $1/16$ of the whole $256\times256$ 
slice ($16<x<32$, $16<y<32$) for four different codes: SLH-MMG,64 
({\it upper left panel}), 
PSPH\,64 ({\it lower left panel}), SLH-P$^3$M\,64 
({\it upper
right panel}), and TVD\,256 ({\it lower right panel}): ({\it a\,}) the density
distribution; ({\it b\,}) the temperature distribution.}

\def\capGT{
The same as Fig.\reffigGD\ except SLH-MMG,64 and  SLH-P$^3$M\,64 are
replaced by SLH-P$^3$M\,64 and SLH-P$^3$M\,128 respectively.}

%
%

\section{Introduction}

Numerical cosmological simulations are now widely used to assess properties
of various cosmological models and to study theoretical aspects of large-scale
structure evolution. The variety of numerical methods employed ranges from high
resolution collisionless $N$-body simulations 
(e.g.\ Davis et al.\markcite{Dea85} 1985; 
Park\markcite{P90} 1990; 
Bertschinger \& Gelb\markcite{BG91} 1991; 
Gelb \& Bertschinger\markcite{GB94a} 1994a,\markcite{GB94b} 1994b; 
Xu\markcite {X95} 1995)
to sophisticated gas dynamics methods 
(e.g.\ Cen \& Ostriker\markcite{CO92a} 1992a; 
Katz, Hernquist, \& Weinberg\markcite{KHW92} 1992; 
Evrard, Summers, \& Davis\markcite{ESD94} 1994; 
Bryan et al.\markcite{Bea94} 1994; 
Summers, Davis, \& Evrard\markcite{SDE95} 1995)
to numerical algorithms that at some level include
galaxies together with collisionless dark matter and intergalactic gas as a 
distinct third component of a simulation 
(Cen \& Ostriker\markcite{CO92b} 1992b,\markcite{CO93a} 
1993a,\markcite{CO93b} 1993b; 
Frenk et al.\markcite{Fel95} 1995; 
Gnedin\markcite{G96a} 1996a,\markcite{G96b} 1996b).

However, not all of these simulation
methods are reliably tested and well understood.
Historically, collisionless $N$-body methods have been subjected to the most
detailed study and testing. The theory of $N$-body methods is well
developed and their errors and limitations are well understood 
(Hockney \& Eastwood\markcite{HE81} 1981; 
Efstathiou et al.\markcite{Eea85} 1985).
Cosmological hydrodynamics methods range from Eulerian hydrodynamic techniques
borrowed from engineering applications 
(Cen et al.\markcite{Cea90} 1990;
Ryu et al.\markcite{Rea93} 1993;
Bryan, Norman, \& Ostriker\markcite{Bea95} 1995)
to quasi-Lagrangian methods that include the
``Smooth Particle Hydrodynamics'' (SPH) method 
(Evrard\markcite{E88} 1988; 
Hernquist \& Katz\markcite{HK89} 1989)
and ``Moving Mesh approach'' (MMA)
(Gnedin\markcite{G95} 1995; 
Pen\markcite{P95b} 1995b). 
Unfortunately,
currently there is no full understanding of errors and specifics of
various hydrodynamic methods. The first comparisons between various 
cosmological
methods 
(Kang et al.\markcite{Kea94} 1994; 
see also Gnedin\markcite{G95} 1995)
demonstrated that differences between various methods can be so
substantial that only a few statistical quantitative 
results are consistent
between different methods. Further work on comparison of various
numerical techniques is therefore required to 
support quantitative results of cosmological hydrodynamical simulations.

Since the Moving Mesh approach is the most recent and therefore least
understood, it requires special attention in comparing it with other
existing numerical techniques. In this paper we concentrate on comparing
an implementation of the Moving Mesh approach called Softened Lagrangian
Hydrodynamics 
(SLH; Gnedin\markcite{G95} 1995)
with existing $N$-body techniques. One possible
advantage of the Moving Mesh approach is that it offers a way to calculate
gravitational forces with high resolution without employing a ``particle''
approach like P$^3$M or TREE. Since the moving mesh represents a 
single-valued coordinate transformation from quasi-Lagrangian space
into real (physical) space, the Poisson equation in real space can be
reduced to an elliptic partial differential 
equation in quasi-Lagrangian space, which can
then be solved by the standard numerical techniques. In the SLH method thus 
calculated the gravity 
force has also the virtue of having {\it exactly\,} the same resolution as
the hydrodynamic solver has (we elaborate below what this
actually means). There are also efforts under way to develop
a purely $N$-body version of the Moving Mesh approach 
(Pen\markcite{P95a} 1995a).

The Moving Mesh approach looks very promising since it has 
higher spatial resolution in dense regions
than Eulerian codes for the same amount of computational resources and is
substantially faster than SPH methods for the same spatial resolution. We,
therefore, undertook to investigate its accuracy by comparing SLH with the
P$^3$M gravity solver. 
Our detailed tests however have uncovered two serious problems with the
Moving Mesh gravity solver which have not been discussed in the literature; 
the problems we found may appear in a wide range of algorithms, so we report
them in this paper. The first problem is the discreteness noise (shot noise)
in high density regions. It is well known that the shot noise does not
represent a serious problem in traditional gravity solvers 
(Hernquist, Hut, \& Makino\markcite{HHM93} 1993),
where force errors effectively average out in high density regions;
however, in the Moving mesh approach the
relationship between the density and the gravitational force is nonlinear,
with the nonlinearity coming from the nonlinear
dependence of the volume of a mesh element on the force acting on it.
In this case the shot noise does not necessarily average out as in the linear
case, leading to spurious numerical heating of high density regions. Secondly,
a solution to the finite-difference realization of the Poisson equation on 
a strongly deformed mesh does not necessarily converge to the true solution,
giving the incorrect force law. 
We propose a solution to the first problem in \S2, but can not resolve the
second one. We have therefore been forced to abandon the Moving Mesh gravity
solver and 
have turned instead to
a P$^3$M gravity solver that is known to work correctly and accurately.
However, while developing the new SLH-P$^3$M code, we encountered another
problem which is generic for (and potentially present in) any self-gravitating 
hydrodynamic code, not only
a Moving Mesh code: since the gravity force solver and the hydrodynamic solver
may treat spatial gradients differently, they are not a priori consistent
with each other.
This inconsistency may
lead to a serious local nonconservation of energy even if, globally, energy is
well conserved. We discuss this effect
in detail in \S3, where we also derive the ``Consistency Condition'' that
every cosmological hydrodynamic code must obey in order to be energy
conserving and apply it in \S4 to develop a new SLH-P$^3$M code. Finally,
we repeat  some of the 
Kang et al.\markcite{Kea94} (1994)
tests for the new code in \S5. We summarize our conclusions in \S6.

\section{Failure of Moving Mesh Gravity}

The Softened Lagrangian Hydrodynamics method was extensively tested
in Gnedin\markcite{G95} (1995);
however, the main emphasis there was on the hydrodynamic
properties of the code. Later, we proceeded to test the gravitational
solver of the SLH code and discovered interesting problems with the
Moving Mesh gravity solver and potentially with self-gravitating 
hydrodynamic solvers
in general. The purpose of this paper is to report the results of those
tests and our solutions to the problems.

We have chosen a CDM+$\Lambda$ model as a framework for our tests. The 
cosmological parameters we use are as follows: $\Omega_0=0.4$,
$h=0.65$, and $\sigma_8=0.85$. We use the BBKS transfer function
(Bardeen et al.\markcite{BBKS86} 1986)
to set up initial conditions. The model is {\it COBE\,}-normalized and is
in a reasonable agreement with all existing large- and 
intermediate-scale observations 
(Kofman, Gnedin, \& Bahcall\markcite{KGB93} 1993; 
Peacock \& Dodds\markcite{PD94} 1994; 
Ostriker \& Steinhardt\markcite{OS95} 1995).
We have run several simulations with $32^3$ and $64^3$ grids all of
them having the initial (Eulerian) comoving 
cell size (or, equivalently, the mean
interparticle separation) equal $0.5h^{-1}\dim{Mpc}$, and, thus, a
$32^3$ simulation has a $16h^{-1}\dim{Mpc}$ box size and a $64^3$
simulation has a $32h^{-1}\dim{Mpc}$ box size. We set both the P$^3$M and
SLH softening parameter to $1/10$ (the softening length is $50h^{-1}\dim{kpc}$)
and keep it constant throughout
our tests. Our simulations have box sizes too small to be of interest
for large-scale structure formation at the current epoch, but they have
sufficient resolution to push simulations into the highly nonlinear regime
to test the SLH method in challenging conditions. In addition, we assume
that the gas is ``adiabatic'', i.e.\ no radiative processes like cooling or
ionization are taken into account. Our current goal is to investigate in 
detail gravitational properties of the SLH code and, therefore, we do not
need to complicate our tests with radiative physics.

\placefig{
\begin{figure}
\insertfigure{\figdir/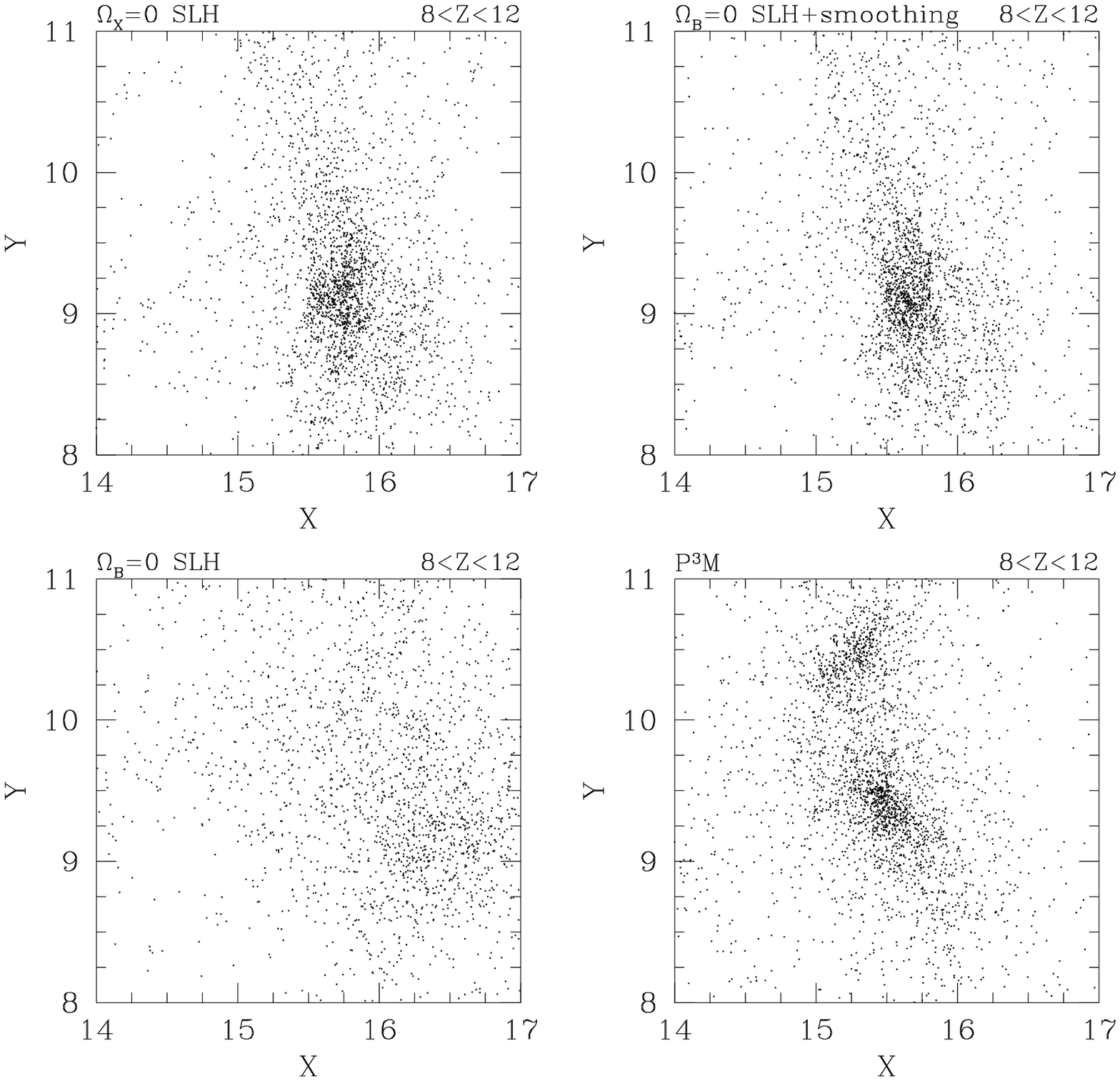}
\caption{\label{figSF}\capSF}
\end{figure}
}

We mostly concentrate on properties of the largest clumps found in our
simulations,
unlike Kang et al.\markcite{Kea94} (1994), who mainly concentrated on 
some average statistical properties of gas simulations with different sizes
\footnote{However, we repeat some of the tests of Kang et al.\markcite{Kea94}
(1994) in \S5.}.
First, we have run SLH and P$^3$M $32^3$ simulations with identical initial
conditions and with $\Omega_b=0$ for the SLH code (the SLH code is thus used
as a collisionless $N$-body solver, however, the hydrodynamic part is used
to follow the mesh deformation as the mesh is tied to baryonic gas in the
SLH method). Fig.\ref{figSF} shows dark matter particles in a subvolume of
these simulations containing the largest clump at current epoch. 
The lower left panel shows particle positions for a SLH 
simulation, and the lower right position shows P$^3$M particles. One can easily
see that the SLH method fails to produce a clump nearly as dense as the P$^3$M
clump. In our comparison we explicitly assume that the P$^3$M gives results
that are correct for dark matter ($\Omega_b=0$) up 
to the softening length and we use it as a template against which to
compare the SLH results. 

For comparison, we have also run the same SLH simulation but with 
$\Omega_b=\Omega_0$, which implies that the dark matter does not gravitate.
The corresponding subvolume is plotted in Fig.\ref{figSF}, upper left panel,
and, as can be  easily seen, there is much better agreement between the
SLH and P$^3$M simulations in that case
\footnote{The SLH gas feels isotropic
pressure while the dark matter has an anisotropic pressure tensor, so some
differences are expected.}.
We must therefore ask ourselves what
is the difference between the SLH $\Omega_b=0$ and SLH $\Omega_X=0$ 
cases that
causes such a significant deviation. 

There are two principal 
differences between dark matter only and gas only simulations with SLH. 
First, the dark
matter density assignment uses the Cloud-In-Cell method to assign the dark
matter density on a quasi-Lagrangian mesh (as explained below), and this
assignment suffers from discreteness noise (or, in other words, shot noise).
Second, the baryonic pressure tensor is isotropic, whereas the dark matter
pressure tensor is not. However, the second reason can not explain why a 
cluster failed to form with $\Omega_b=0$ SLH, 
since in the P$^3$M simulation the dark matter has 
anisotropic pressure tensor as well, but a dense cluster forms.
We conclude therefore that it is the SLH density assignment scheme that is
responsible for the cluster diffusion problem.
 
In order to elaborate on this effect, we recall 
the basic
ingredients of the Moving Mesh approach in general and the SLH method in
particular (the reader should refer 
to Gnedin\markcite{G95} (1995) for complete explanation
of notation and terminology). 
Let us consider a general coordinate transformation, which connects the real
space coordinates, $x^i$, with some other coordinate system $q^k$, which we 
will call ``quasi-Lagrangian'' hereafter,
\begin{equation}
        x^i = x^i(t,q^k).
  \label{xasfunq}
\end{equation}
There are no restrictions on $q^k$ up to now, but we will assume in the 
following that the new coordinate system $q^k$ has the property that it
redistributes numerical resolution into high density regions. The important
quantity that controls the degree of deformation between coordinate systems
$x^i$ and $q^k$ is the {\it deformation tensor\,},
\begin{equation}
        A^i_k \equiv {\partial x^i\over\partial q^k}.
        \label{DeformationTensor}
\end{equation}
Then the density of gas or dark matter can be represented as
\begin{equation}
	\rho = \rho_0/\A,
	\label{density}
\end{equation}
where $\A$ is the determinant of the deformation tensor $A^i_k$,
\begin{equation}
	\A \equiv \det{A^i_k},
	\label{detA}
\end{equation}
and $\rho_0\equiv\rho_0(q^k)$ 
is the quasi-Lagrangian density,
or, in other words, the mass of a fluid element having a unit volume in
quasi-Lagrangian space $q^k$. Note that quasi-Lagrangian coordinates may
deviate significantly from exactly Lagrangian ones; i.e., for a given mass 
element $q^k$ need not be constant.

In a Moving Mesh approach (and, therefore, in the SLH code as well), 
coordinates $x^i$ are real space images of vertices of a uniform mesh in
quasi-Lagrangian space, $q^1_{ijk} = i$, $q^2_{ijk} = j$, $q^3_{ijk}=k$.
Since $\A$, as a function of the deformation tensor,
is the property of the mesh, it is a smooth function of position
(as smooth as the actual mesh equation allows it to be). For baryonic gas,
$\rho_{0,B}$ is also a smooth function of position since it represents the
mass of the gas in a cell; for a fully Lagrangian flow $\rho_{0,B}\equiv1$.
However, this is not necessarily true for the dark matter density, which may,
therefore, have large gradients in $q$-space.
In the SLH
approach the dark matter density that sources the right hand side of the 
Poisson equation is determined by assigning dark matter mass to cells in the
quasi-Lagrangian space with the cloud-in-cell technique:
\begin{equation}
	\rho_{0,X}(q^l) = \sum_{i,j,k} m_{i,j,k} W(q^l-q^l_{i,j,k}).
	\label{dmdenCIC}
\end{equation} 
In voids where, by construction, there are always almost the same number of
dark matter particles per cell (since in voids the baryonic flow is 
Lagrangian and the pressure is negligible, dark matter velocity is equal to 
the gas velocity at the same position and, therefore, both the dark matter
and baryons move simultaneously in voids), and, therefore, $\rho_{0,X}$ is
a smooth function of position and is almost constant. (It would be exactly
constant before dark matter shell crossing and before gas shocking if the
pressure in voids was zero and quasi-Lagrangian space was exactly Lagrangian.)
This is not so  in high density regions. Dark matter particles wander
around in a cluster while the mesh does not move substantially. 

In Eulerian
PM codes the density fluctuations are a less severe (while still existing) 
problem since there are many dark matter 
particles per one PM cell in a high density region 
and the dark matter density is determined accurately
with the Cloud-In-Cell method. However, in the Moving Mesh approach,
if the number of dark matter particles is equal to the number of baryonic 
cells, there is on average one (or slightly more than one
if the dark matter is more concentrated than the baryons) dark matter 
particle per cell even in the highest density region since the high density
is achieved not by collecting many particles in one immobile cell but by
shrinking a cell to small volume keeping its mass approximately constant.
If these particles are distributed randomly with respect to the mesh, 
i.e.\ if all $q^l_{i,j,k}$ are random positions, 
equation (\ref{dmdenCIC}) would predict that $\rho_{0,X}$ fluctuates
on a scale of one cell with the amplitude comparable with its mean value.
That means that the density $\rho_X$ in clusters would fluctuate with
the amplitude comparable with its mean value even in the cluster center.
Since particles move, the dark matter density will fluctuate both spatially
and temporally. The temporal fluctuations, however, seem to be less serious
since the gravitational time scale in clusters is usually larger than the
hydrodynamic Courant time scale.

Such enormous fluctuations will produce fluctuations in the gravitational 
potential; it has been shown 
(Hernquist, Hut, \& Makino\markcite{HHM93} 1993)
that those fluctuations effectively
average out for traditional gravity solvers. However, in the Moving Mesh 
approach the situation is different. Since the mesh is usually driven
by gravity force, and the gravity force itself is calculated on the mesh,
there is a nonlinear dynamical relation between the gravitational force and,
say, the density defined on the mesh (with nonlinearity coming from the
nonlinear dependence of the deformation tensor on the gravitational force).
Therefore, the traditional analysis 
(Hernquist, Hut, \& Makino\markcite{HHM93} 1993),
where the relation between
the force and the density is linear, does not apply and force 
errors do not necessarily average out as in the linear case. The nonlinearity
between the density and the force also serves as a dissipative
numerical mechanism, which leads to eventual blowing up of a cluster. 
We would like
to stress here that this effect is a direct consequence of the shot noise
effect in clusters, when in the Moving Mesh approach there are only a few
dark matter particles per cell even in the highest density regions.

\placefig{
\begin{figure}
\insertfigure{\figdir/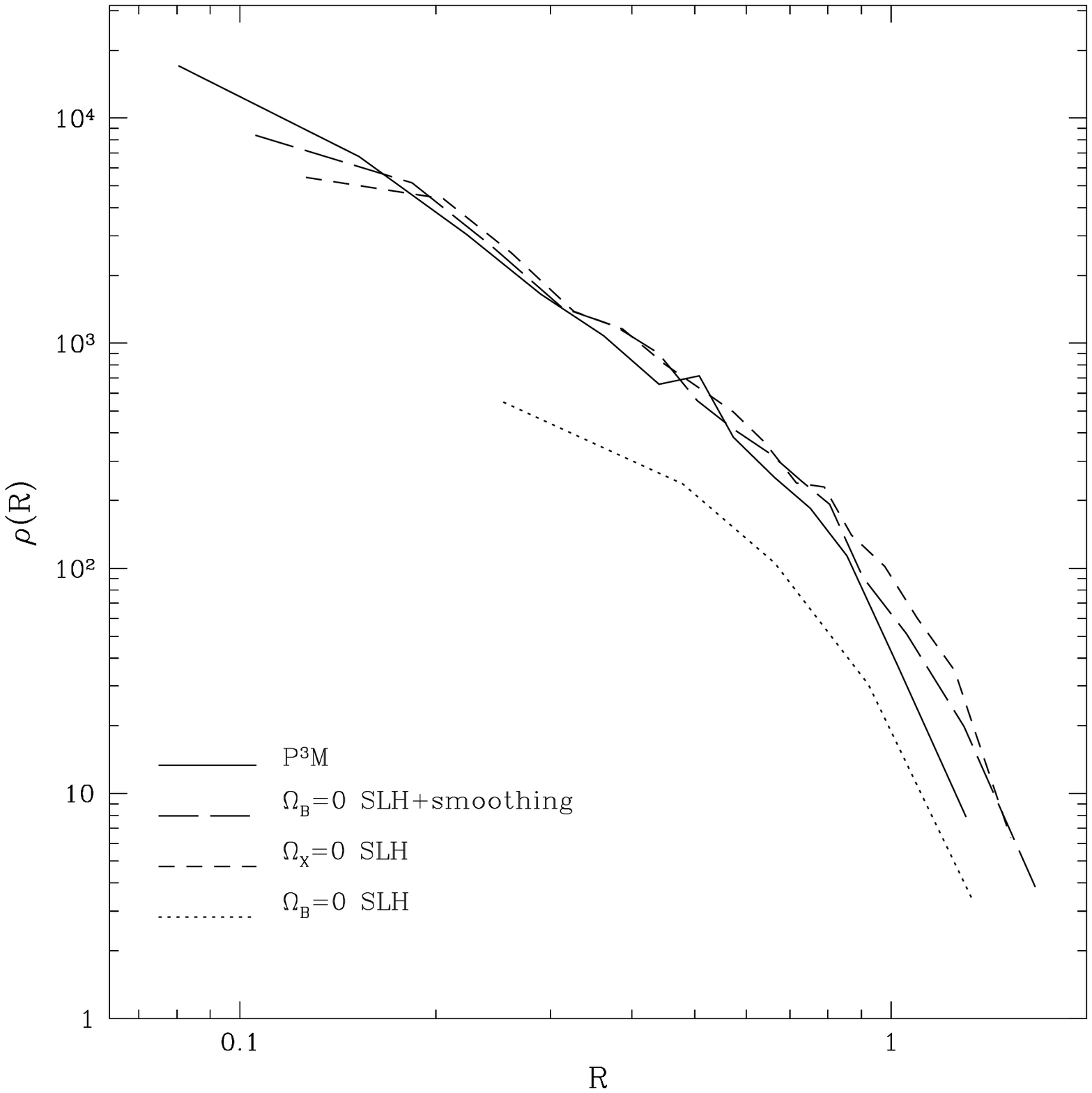}
\caption{\label{figSP}\capSP}
\end{figure}
}

It is easy to demonstrate that this shot noise 
effect is indeed present in Moving Mesh
gravity solvers. First, we noticed above,
that this effect is absent when the baryons instead of the dark matter is
gravitationally dominant, and $\rho_{0,B}$ is always a smooth function of
position. Second, we can define $\rho_{0,X}$ as a smooth function of position
if we smooth it in quasi-Lagrangian space. This procedure will not 
significantly degrade
the resolution since it is $\A$, not $\rho_{0,X}$, that carries the high
resolution. The smoothing will not affect voids since $\rho_{0,X}$ is
almost constant in voids; it will affect places where the dark matter
density is physically different from the baryon density, i.e.\ near shocks.
We can hope however that these regions occupy a small fraction of the total
volume, and, what is more important, they do not dominate the gravitational
potential. 

To test the shot noise hypothesis we have run another 
SLH simulation where we have
smoothed the quasi-Lagrangian dark matter density $\rho_{0,X}$ entering 
the right hand side of the Poisson 
equation with a Gaussian filter with width equal to three 
cell sizes in quasi-Lagrangian space (we would like to stress that this
is {\it not\,} equivalent to smoothing the dark matter density, since the
factor $\A$ remains unsmoothed). The resulting particle positions are 
plotted in Fig.\ref{figSF}, upper right panel. One can easily see that a 
cluster is now formed. Fig.\ref{figSP} shows density profiles for
the clusters in these four cases after the cluster was identified with the
DENMAX algorithm 
(Bertschinger \& Gelb\markcite{BG91} 1991).
The SLH gravity with smoothing 
is perhaps a factor of 1.5 softer than the Plummer law for these simulations
\footnote{We have also performed the same SLH simulation without smoothing
but with 8 and 64 times more dark matter particles; there is a definite
improvement as the number of dark matter particles increases, but the rate
of improvement is very slow, and even with
64 times more dark matter particles than baryonic cells the cluster under
consideration is still significantly diffused out.}.

\placefig{
\begin{figure}
\insertfigure{\figdir/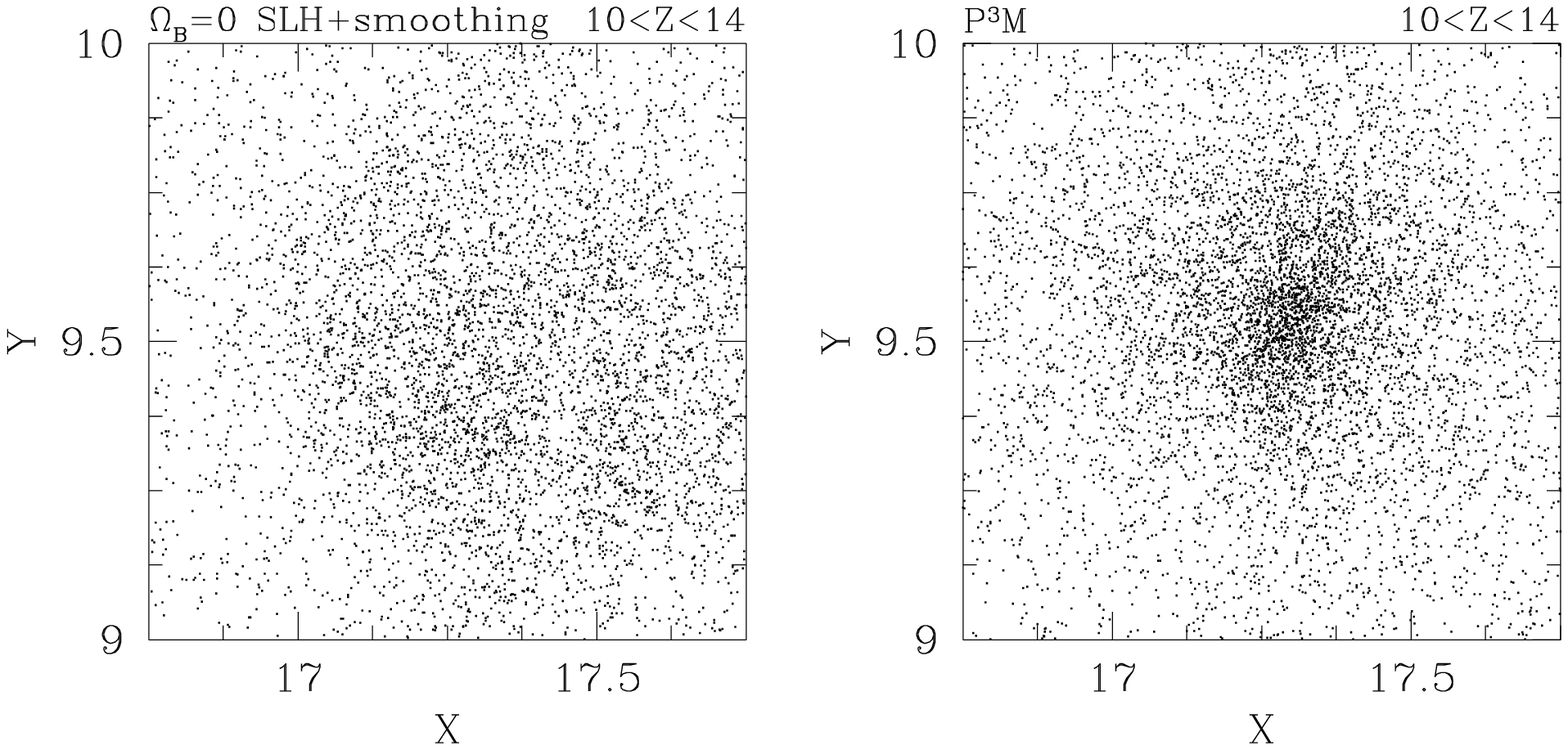}
\caption{\label{figST}\capST}
\end{figure}
}

In order to test the Moving Mesh gravity further, we have performed two
$64^3$ simulations with the parameters described above:  one is a P$^3$M 
simulation and the other one is a SLH simulation with $\Omega_b=0$ and
dark matter smoothing as explained in the previous paragraph. Particles
around the heaviest cluster in the two simulations 
are plotted in Fig.\ref{figST}
in the right and left hand panels respectively. One can see that at
higher resolution ($64^3$ vs $32^3$ in Fig.\ref{figSF}) 
the SLH cluster again diffuses outwards. Fig.\ref{figSQ}
shows the density profile for the cluster. Obviously, the SLH gravity
effectively softens at a scale a factor of 6 or 7 larger than the
formal softening length. Since the effect of shot noise in high density
regions is eliminated in the SLH simulation, one has to admit the
existence of yet another problem with the SLH gravity. 

\placefig{
\begin{figure}
\insertfigure{\figdir/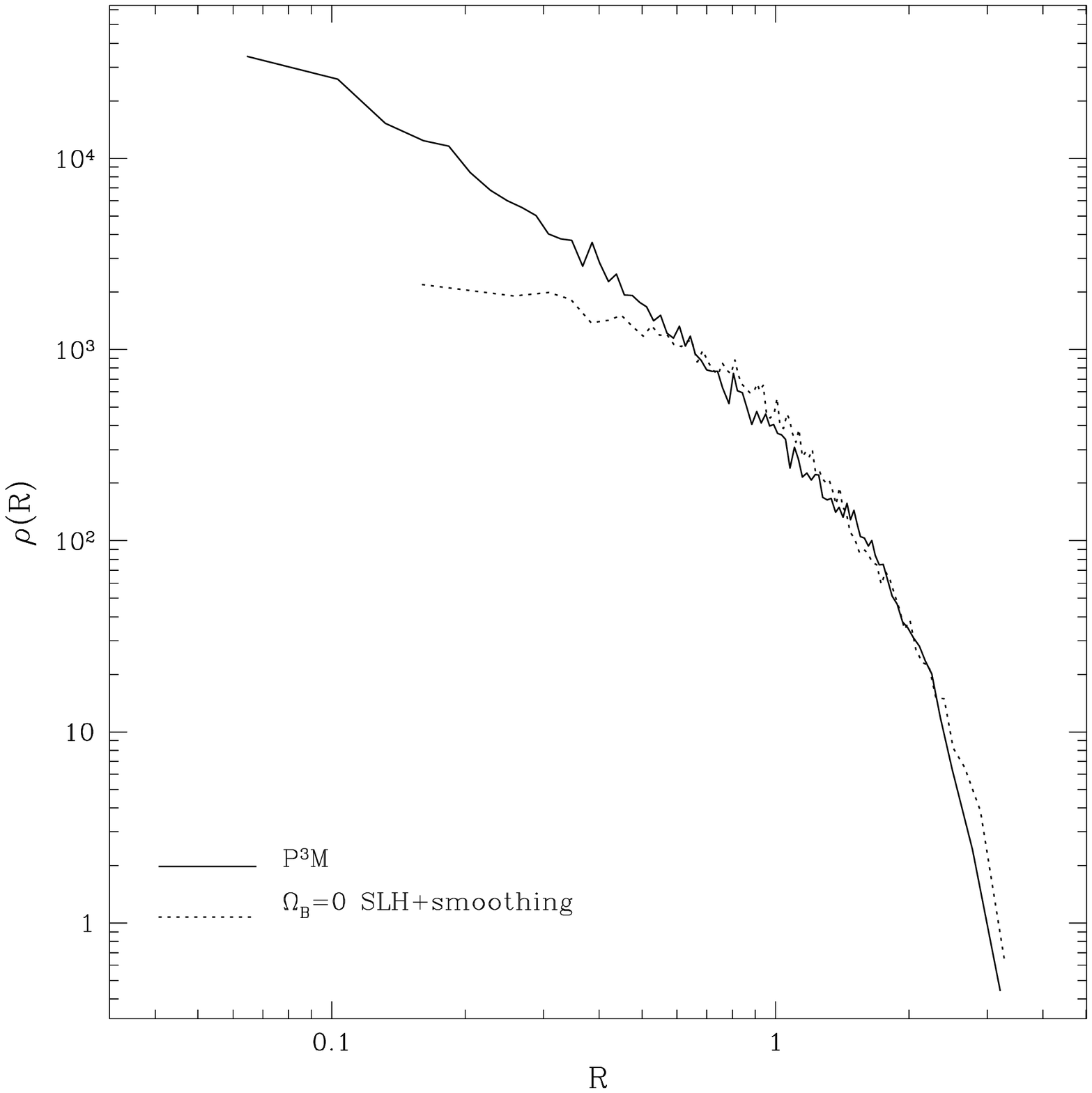}
\caption{\label{figSQ}\capSQ}
\end{figure}
}

In order to 
investigate the accuracy of the SLH Poisson solver we calculated point
mass gravitational forces for ten randomly chosen positions on the SLH
mesh taken from the $64^3$ SLH simulation at $z=0$.
The corresponding force laws are plotted in Fig.\ref{figGF}, left panel.
Since points are chosen randomly on a mesh, their effective smoothing
lengths are different (for example in a void gravitational smoothing is
larger than the original cell size since a cell in a void has expanded
during the evolution); in general, however, there is an acceptable agreement
between the SLH point mass force and $1/R^2$ law (dotted line in 
Fig.\ref{figGF}).
Now, instead of ten randomly chosen points, we picked up two points that lay
close to the center of the biggest cluster in the SLH simulation. Their point
mass force laws are plotted in Fig.\ref{figGF} with solid lines. These laws
deviate significantly from the $1/r^2$ law at 5 to 6 softening lengths even
if these cells have sizes of the order of a softening length. 

\placefig{
\begin{figure}
\epsscale{0.7}
\insertfigure{\figdir/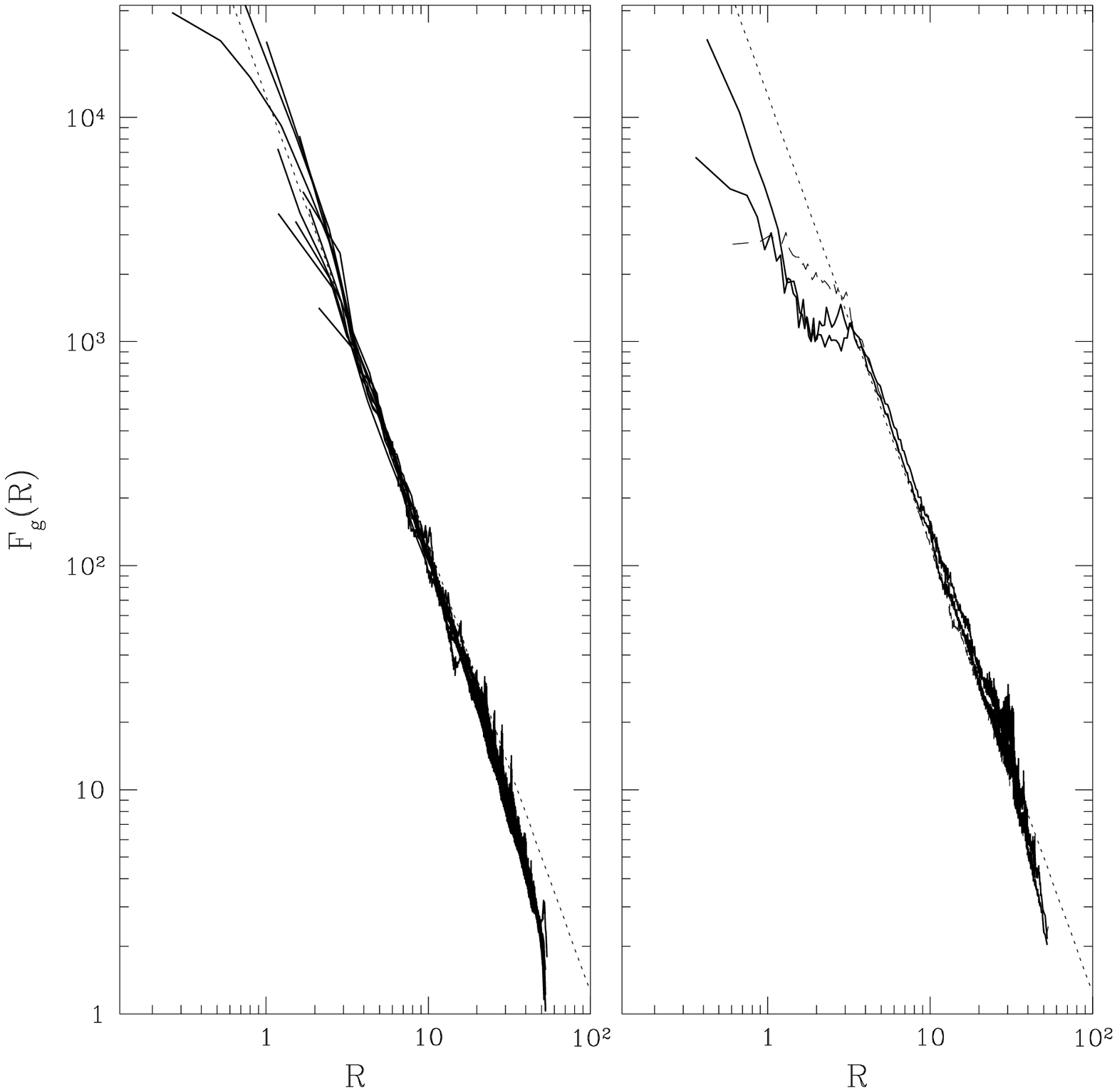}
\caption{\label{figGF}\capGF}
\end{figure}
}

What causes these errors in gravity in a cluster? In order to test if 
it is due to the loss of accuracy in the finite differencing of
the gravitational potential, 
instead of solving the Poisson equation
\begin{equation}
	\Delta \phi = {3 \Omega_0 a\over2} \delta,
\end{equation}
we solved three Poisson equations for each of force component separately,
\begin{equation}
	\Delta f_i = {3 \Omega_0 a\over2} {\partial \delta\over\partial x^i}
\end{equation}
which eliminates the need to differentiate the potential but requires
at least three times the CPU time used by the potential solver.
The corresponding point mass force law is plotted with the thin dashed line 
in the
right panel of Fig.\ref{figGF}. While there is some improvement in solving
for the force compared to solving for the potential, it is obviously
insufficient to improve the Moving Mesh gravity solver. We specifically stress
here that in this test, the specifics of the SLH algorithm
come only with the
strongly deformed mesh; the Poisson solver is a {\it generic\/} feature
of any Moving Mesh approach and it is this that fails. We, thus, have
encountered another problem with the Moving Mesh gravity: when the mesh
is strongly deformed, the Poisson solver occasionally gives incorrect
results. It seems quite difficult to find out what specifically is wrong
with the Poisson solver or what properties of the mesh cause such a big
error. It is also not clear if the less deformed mesh has the same problem
but less pronounced so that it avoids clear recognition.

Even if the problem we have highlighted appears only in small volume of
the whole simulation, there is no guarantee that the rest of a simulation
is correct or has a specified resolution. Until the specific problem with
the Moving Mesh gravity is identified and resolved, we feel 
that it is unsafe to
use the Moving Mesh gravity solver for cosmological simulations. The full
analysis of Poisson solver errors on strongly deformed meshes is
beyond the frame of this
paper and, therefore, we have decided to abandon the Moving Mesh gravity
solver and proceed into combining the SLH hydrodynamic method with the
P$^3$M solver which is known to give correct results
(Bertschinger\markcite{B91} 1991). We shall see that
merging these codes introduces another problem to solve.

\section{Consistency Condition}

We now imagine building a cosmological hydrodynamic code by combining
separate hydrodynamic and gravity solvers together. Both methods can
calculate hydrodynamic and gravitational forces on a resolution element
scale quite differently, and there is a priori no guarantee that these
two methods would be consistent with each other.

Let us consider an example. Let us imagine that we constructed an
isothermal gas sphere; we utilize so many resolution elements
(cells or particles), that we do not have to worry about relaxation
processes. We now insert this object into a hydro+gravity code which
has gravitational force softening much shorter than the pressure force
softening\footnote{This should be considered only as a thought experiment;
in reality hydrodynamic codes have pressure softening comparable to
the resolution element size and, thus, our assumption of having many
resolution elements per pressure or gravity softening length is unrealistic.
To implement it in practice, one would require to smooth both pressure
and gravity force over a scale encompassing many resolution elements.}.
The object, which would be in real 
physical equilibrium (i.e.\ in equilibrium with exact force laws for both
gravity and pressure force), would collapse further since the softened
gravity force would be stronger than the softened pressure force due to
larger pressure force softening. The object would thus form a small and
dense clump that is a purely numerical artifact. This process is an
example of a more general inconsistency between hydro and gravity solvers
that we discuss in detail below.

Let us consider the hydrodynamic equations with self-gravity in a general
coordinate system $q^k$ as defined above
(at this point we do not require that the coordinate system $q^k$ be close
to Lagrangian; in particular, it may be fully Eulerian when $x^i=q^i$).
Let $\rho$ be the density, $v^i$ be the velocity, $P$ be the pressure,
$E\equiv\rho v_i^2/2 + P/(\gamma-1)$ be the total energy density, and $f_i$
be the gravitational force per unit mass of the fluid with 
the polytropic index $\gamma$.
The Euler equations in a coordinate system $q^k$ read:
\begin{mathletters}
\begin{equation}
        {\partial\rho_0\over\partial t} + 
        {\partial\over\partial q^k}(\rho_0 w^k) = 0,
        \label{MIXEDd}
\end{equation}
\begin{equation}
        {\partial\rho_0 v_i\over\partial t} + {\partial\over\partial q^k}
        \left(\rho_0 v_i w^k  + \A B^k_i P\right) =  
        \rho_0 f_i,
        \label{MIXEDm}
\end{equation}
and
\begin{equation}
        {\partial E_0\over\partial t} + {\partial\over\partial q^k}
        \left(E_0 w^k + v^i\A B^k_i P\right) =
        \rho_0 v^i f_i,
        \label{MIXEDe}
\end{equation}
where
\label{MIXED}
\end{mathletters}
the deformation tensor $A^i_k$ has been defined above,
$B^i_k$ is the matrix inverse to $A^i_k$,
$$
        B^i_jA^j_k = A^i_jB^j_k = \delta^i_k,
$$
$\A\equiv\mbox{det}(A^i_k)$ is the determinant of the deformation 
tensor, which is equal to the volume in real space of a fluid element with
the unit volume in quasi-Lagrangian space $q^k$, $d^3x=\A d^3q$,
$\rho_0$ is the quasi-Lagrangian mass
density,
\begin{equation}
        \rho_0 \equiv \rho\A,
\end{equation}
$E_0$ is the quasi-Lagrangian total energy density,
\begin{equation}
        E_0 \equiv E\A,
\end{equation}
and 
\begin{equation}
        w^k \equiv B^k_i(v^i-\dot{x}^i),
\end{equation}
with the dot standing for the partial time derivative, $\dot{x}^i(t,q^k)=
\partial x^i(t,q^k)/\partial t$.

If we now consider a hydrodynamic code that solves these equations, we
must interpret the spatial derivative $\partial/\partial q^k$ as a 
numerical derivative (for example, a finite difference derivative for
a mesh-based code), and we denote it as $\partial_N/\partial q^k$,
and we must include numerical dissipation terms
responsible for shock handling and stability. Let us introduce a numerical 
density flux correction ${\cal F}^k$, a numerical momentum flux correction
${\cal G}^k_i$, a numerical total energy flux correction ${\cal H}^k$, and
a numerical gravitational energy correction $\Lambda$
through the following formulas:
\begin{mathletters}
\begin{equation}
        {\partial\rho_0\over\partial t} + 
        {\partial_N\over\partial q^k}(\rho_0 w^k+{\cal F}^k) = 0,
        \label{MIXEDdnum}
\end{equation}
\begin{equation}
        {\partial\rho_0 v_i\over\partial t} + {\partial_N\over\partial q^k}
        \left(\rho_0 v_i w^k  + \A B^k_i P + {\cal G}^k_i
        \right) = \rho_0 f_i,
        \label{MIXEDmnum}
\end{equation}
and
\begin{equation}
        {\partial E_0\over\partial t} + {\partial_N\over\partial q^k}
        \left(E_0 w^k + v^i\A B^k_i P + {\cal H}^k
        \right) = \rho_0 v^i f_i + \Lambda.
        \label{MIXEDenum}
\end{equation}
Among those four, only ${\cal F}^k$ and $\Lambda$ will be of interest
to us.
\label{MIXEDnum}
\end{mathletters}

We now want to establish the total energy conservation. The total energy
can be represented as the sum of the total gas energy,
$$
	{\cal E} = \int_N E_0 d^3q,
$$
and the gravitational energy,
$$
	{\cal W} = {1\over2} \int_N \rho_0 \phi\, d^3q,
$$
where we understand integrals in a numerical sense and denote this by
index $N$, and $\phi$ is a gravitational potential, and is 
determined by the following
equation:
\begin{equation}
	\phi(q^k) = \int_N G(|x^i(q^k)-x^i(q^k_1)|) \rho_0(q^k_1) d^3q_1,
	\label{phi}
\end{equation}
and $G(r)$ is a Green function for a gravity solver. It is important to note
here that this function is arbitrary, and, therefore, our consideration
is equally applicable to a softened gravity law as well as to the $1/r^2$ law.
Moreover, we do not need to specify a softening length for the gravity.
It will be shown below that the problem of inconsistency does not reduce to the
problem of different resolution scales for gravity and gas and is more
generic.

We now demand (neglecting cosmological expansion for the moment) that 
\begin{equation}
	{d\over dt}({\cal E}+{\cal W}) = 0.
	\label{Energyconserv}
\end{equation}
It can be easily seen that
\begin{equation}
	{d\over dt} {\cal E} = \int_N {\partial E_0\over\partial t} d^3q =
	 \int_N \left(\rho_0 v^i f_i + \Lambda\right) d^3q
	\label{dEdt}
\end{equation}
provided the numerical integral of a numerical divergence is zero (we assume
that this is the case).
For the time derivative of the gravitational energy, we obtain:
\begin{equation}
	{d\over dt} {\cal W} = \int_N {\partial\rho_0\over\partial t} \phi\,
	d^3q + \int_N \rho_0 \dot{x}^i {\partial\over\partial x^i} \left[
	\int_N 
	G(|x^i-x^i(q^k_1)|) 
	\rho_0(q^k_1) d^3q_1 \right] d^3q.
	\label{dWdt}
\end{equation}
It is important to note that the derivative with respect to $x^i$ that enters
the last equation is a {\it full derivative\,}, not the numerical one. We can
introduce the new quantity, $\hat{f}_i$, defined as:
\begin{equation}
	\hat{f}_i \equiv -{\partial\phi\over\partial x^i}  =
	-{\partial\over\partial x^i} \left[
	\int_N G(|x^i-x^i(q^k_1)|) \rho_0(q^k_1) d^3q_1 \right] d^3q,
	\label{hatfidef}
\end{equation}
which is the true gradient (with the minus sign) 
of a numerical gravitational potential,
i.e.\ the gradient which will be computed in a simulation,
in which all spatial gradients are computed mathematically exactly
(for example, a P$^3$M code possesses this property).
Note, that due to the finite number
of resolution elements in a real simulation, $f_i$ is not
necessarily equal to $\hat{f}_i$, but rather is a numerical
approximation to it.
Equation (\ref{dWdt}) can now be written as follows:
\begin{equation}
	{d\over dt} {\cal W} = \int_N {\partial\rho_0\over\partial t} \phi\,
	d^3q - \int_N \rho_0 \dot{x}^i \hat{f}_i d^3q.
	\label{dWdtb}
\end{equation}

We now assume that the numerical derivative $\partial_N$ 
obeys the product rule, i.e.\
\begin{equation}
	{\partial_N\over\partial q^k}(fg) = f {\partial_N g\over\partial
	q^k} + g {\partial_N f\over\partial q^k}.
	\label{productrule}
\end{equation}
This is true for dimensionally split mesh codes with central
differences and can be true in SPH
codes with appropriate choice of weightening scheme; it is generally not true
in the SLH code since it is not dimensionally split, and, therefore, the
SLH code has additional energy errors compared to dimensionally split
mesh codes. (These errors are typically small and random and seem not to
contribute significantly to the total energy error, although, in principle,
nothing prevents them from causing a significant energy error.)

Then we can use equation (\ref{MIXEDdnum}) to reduce equation (\ref{dWdtb})
to the following expression:
\begin{equation}
	{d\over dt} {\cal W} = \int_N (\rho_0 w^k+{\cal F}^k)
        {\partial_N\phi\over\partial q^k} d^3q
	- \int_N \rho_0 \dot{x}^i \hat{f}_i d^3q.
	\label{dWdtc}
\end{equation}
Combining both equation (\ref{dEdt}) and equation (\ref{dWdtc}) with
equation (\ref{Energyconserv}), we finally obtain:
\begin{equation}
	\int_N \left( \Lambda + {\cal F}^k
        {\partial_N\phi\over\partial q^k}\right) d^3q +
	\int_N \rho_0 \left[ v^i f_i - \dot{x}^i \hat{f}_i +
	(v^i-\dot{x}^i)B^k_i{\partial_N\phi\over\partial q^k}\right]
	 d^3q = 0.
	\label{consista}
\end{equation}

We can immediately derive the form for $\Lambda$ from the equation 
(\ref{consista}) taking into account that $\Lambda$ is the numerical
correction to the gravitational energy change
and it must vanish when ${\cal F}^k$ vanishes:
\begin{equation}
	\Lambda = - {\cal F}^k {\partial_N\phi\over\partial q^k}.
	\label{Lamdadef}
\end{equation}
Now we are left with the following equation, which must be
satisfied if the energy is to be conserved:
\begin{equation}
	\int_N \rho_0 \left[ v^i f_i - \dot{x}^i \hat{f}_i +
	(v^i-\dot{x}^i)B^k_i{\partial_N\phi\over\partial q^k}\right]
	 d^3q = 0.
	\label{consistb}
\end{equation}
Let us examine this carefully. There are three gravitational forces that
enter this equation: the original force that a full gravity solver provides,
$\hat{f_i}$, the force that the hydrodynamic code feels, $f_i$, and the
numerical gradient of the gravitational potential, $B^k_i\partial_N\phi/
\partial q^k$. The last expression is the gradient of a scalar function
as the hydrodynamic code understands it; for a mesh-based code it is
a finite difference, for a SPH-like code it is the function values weighted
by the gradients of the kernel, etc.

If we require that the self-gravitating hydrodynamic code 
should conserve energy
everywhere and for all possible cases, we must satisfy the 
{\it Gravitational Consistency
Condition\,} at every resolution element:
\begin{equation}
	v^i f_i = \dot{x}^i \hat{f}_i -
	(v^i-\dot{x}^i)B^k_i{\partial_N\phi\over\partial q^k}.
	\label{consist}
\end{equation}
Strictly speaking, we may add a divergence term,
which always can be hidden into ${\cal H}^k$. This condition should be 
considered as a restriction that the gravitational force $f_i$ should satisfy
in order for the self-gravitating hydrodynamic code to be energy conserving.

It is instructive to consider some special cases:
\begin{enumerate}
\item A fully Lagrangian code (in particular, SPH). For a fully Lagrangian
      code $\dot{x}^i=v^i$ and equation (\ref{consist}) is satisfied if
      $f_i=\hat{f}_i$. Thus, any gravitational solver is consistent with
      the fully Lagrangian code; in other words, a fully Lagrangian code
      is a priori consistent with a gravity solver and no special care should
      be taken to satisfy the Consistency Condition.
\item A fixed mesh finite-difference code (in particular, a 
      fully Eulerian finite-difference code). 
      For this case, we can assume $\dot{x^i}=0$ and
      the Consistency Condition is satisfied if
$$
	f_i = -B^k_i{\partial_N\phi\over\partial q^k}.
$$
      We conclude, therefore, that it is essential that the gravitational 
      force in a fixed mesh code be a finite difference gradient of the
      gravitational potential. In particular, in a PM + Eulerian 
      finite-difference code 
      combination it is the gravitational potential that should be 
      calculated in the PM part, and not the force.
\item A Riemann solver. In this case the numerical gradient 
      $\partial_N/\partial q^k$ is actually a Riemann solver; it is not
      straightforward to represent the gravitational force as the
      gradient of a potential in this case; rather it is easier to
      incorporate the gravitational force into the Riemann solver itself as:
$$
  	f_i = \hat{f}_i = -B^k_i{\hat\partial_N\phi\over\partial q^k},
$$
      where notation $\hat\partial_N/\partial q^k$ means the Riemann solver
      with the gravitational force incorporated in it; in particular,
      this is the way the KRONOS code works (Bryan et al.\markcite{Bea95} 
      (1995).
\item A Moving Mesh gravity code 
      (i.e.\ a code that uses the Moving Mesh gravity
      as a gravity solver). Strictly speaking, for this case equation 
      (\ref{phi}) does not hold, but we can still consider it here
      recalling that 
$$
	\hat{f}_i = -B^k_i{\partial_N\phi\over\partial q^k}
$$
      for a Moving Mesh Gravity code. In that case $f_i$ is also equal
      to both $\hat{f}_i$ and $-B^k_i\partial_N\phi/\partial q^k$ and
      the Consistency Condition holds.
\end{enumerate}

Thus, the Consistency Condition is usually satisfied for most of 
currently used
numerical techniques.

Let us consider here a simple example. Let us imagine that we have a
one-dimensional
Eulerian finite-difference hydrodynamic code (so that the product rule
[\ref{productrule}] is always satisfied) that keeps all hydrodynamic
quantities at the centers of equal cells numbered by index
$i=1,2,...,N$ (where $N$ is the number of cells and the periodic boundary
condition is assumed), so that the gas density at the cell $i$ is $\rho_i$,
the energy density is $E_i$ etc. This code is combined with the PM
gravity solver that uses the gas density $\rho_i$
to solve the Poisson equation (using, say, the FFT technique)
\begin{equation}
	\phi_{i+1}+\phi_{i-1}-2\phi_i=4\pi G \rho_i
	\label{ex1poisson}
\end{equation}
(we assume here $\Delta x=1$ for simplicity) to obtain
the gravitational
potential $\phi_i$ at the cell centers, derives the gravitational 
force $f_i$ acting
at the cell center by finite differencing the potential,
\begin{equation}
	f_i = (\phi_{i-1}-\phi_{i+1})/2,
	\label{ex1force}
\end{equation}
and uses it to update the gas
velocity $v_i$. The code constructed this way conserves total
energy exactly (provided the hydrodynamic part conserves the kinetic
plus thermal energy exactly, as is the case for conservative schemes) 
as explained above,
i.e.\ it satisfies the Consistency Condition. Now let us imagine that
one decides to replace the PM gravity solver with the exact gravity solver,
which calculates the exact gravitational potential $\hat\phi$ and force 
$\hat f$ given the
density distribution:
$$
	\rho(x) = \left\{
	\begin{array}{ll}
	\rho_1, & 0<x\le1, \\
	\rho_2, & 1<x\le2, \\
	... & \\
	\rho_i, & i-1<x\le i, \\
	... & \\
	\rho_N, & N-1<x\le N \\
	\end{array} \right. 
$$
(One can achieve this, for example, by using the PM gravity solver 
with a much finer mesh.)
If one now uses the exact gravitational force $\hat f_i$ to update
the gas velocity $v_i$, the Consistency Condition will not be satisfied,
and local energy errors will be introduced. However, if one uses the
exact gravitational potential $\hat\phi$ and calculates the gravitational
force $f$ using (\ref{ex1force}) with the new potential, 
the energy conservation
will be restored even if the exact gravitational potential $\hat\phi$ is
different from the potential $\phi$ obtained by solving the Poisson
equation on the hydrodynamic mesh. Thus, the Consistency Condition requires a
consistent treatment of the gravitational potential and forces in the
dark matter and gas.

We would like to stress here again that the Consistency Condition is not
equivalent to having gravity and pressure force softening commensurate;
the Green function in (\ref{phi}) allows for any value for the gravitational 
softening parameter; more that that, since the shape of the Green function
is not specified, it is not required that the force $f_i$ be actually 
gravitational force; any potential force solver (for example, molecular 
forces) should satisfy the Consistency Condition similar to equation
(\ref{consist}). This comment would imply that the requirement of having
gravitational and pressure softenings commensurate is an {\it additional\/}
requirement a self-gravitational hydrodynamic code must satisfy.

\section{A SLH-P$^3$M Code}

We now turn to constructing a SLH-P$^3$M code, where the gravity force
is calculated by a P$^3$M solver and the mesh velocity
$\dot{x}^i$ is not necessarily equal
to the fluid velocity $v^i$. 
In this case it is impossible to solve equation (\ref{consist}) 
exactly, and only an approximate solution can be proposed.

We start by noting that the SLH mesh equation, i.e.\ the equation that
connects $\dot{x}^i$ and $v^i$ in the SLH approach,
\begin{equation}
        {\partial^2\dot{x}^i\over\partial q_m\partial q^m} =
        {\partial\over\partial q_m}
        \left[\left(\delta^i_j-\sigma^i_j\right)
        {\partial v^j\over\partial q^m}\right],
        \label{xdotlaplace3D}
\end{equation}
is a Galilean invariant
modification of the following equation:
\begin{equation}
        \dot{x}^i = (\delta^i_j-\sigma^i_j) v^j
	\label{xdotdirect3D}
\end{equation}
(here $\sigma^i_j$ is the softening tensor that vanishes in the Lagrangian
limit and approaches $\delta^i_j$ in the Eulerian limit; for full definitions
see Gnedin\markcite{G95} (1995) or Appendix A). 
If we consider $\dot{x}^i$ from 
equation (\ref{xdotdirect3D}) as an
approximation to  $\dot{x}^i$ derived from equation 
(\ref{xdotlaplace3D}) and substitute (\ref{xdotdirect3D}) into
(\ref{consist}), we obtain the following (Galilean invariant) equation:
\begin{equation}
	f_i = \hat{f}_i - \sigma^j_i\left(\hat{f}_j + 
   	B^k_j{\partial_N\phi\over\partial q^k}\right).
	\label{ficonsist}
\end{equation}
This equation will serve as our definition for the gravity force on the
gas in the
SLH-P$^3$M code. Both $\hat{f}_i$ and $\phi$  are obtained by means of
a P$^3$M code and $f_i$ is the gravitational force applied to the
baryonic gas. The forces on the dark matter particles are given instead 
by $\hat f_i$. Tests also show that the following simplified version of
equation (\ref{ficonsist}) is a good choice as well:
\begin{equation}
	f_i = \hat{f}_i - \bar\sigma \left(\hat{f}_i + 
   	B^k_i{\partial_N\phi\over\partial q^k}\right),
	\label{ficonsistmax}
\end{equation}
where $\bar\sigma$ is the maximum eigenvalue of the softening tensor
$\sigma^j_i$. One can note that, again,
in the fully Lagrangian ($\bar\sigma=0$) case $f_i = \hat{f}_i$, while 
in the fully Eulerian ($\bar\sigma=1$) regime $f_i = 
-B^k_i{\partial_N\phi/\partial q^k}$.

Equations (\ref{ficonsist}) and (\ref{ficonsistmax}) have a remarkable
feature that they {\it guarantee the commensurate softenings for both
gravity and pressure\/} for the SLH code, since the SLH code switches
from Lagrangian to Eulerian description as density increases and,
therefore, the gravitational force at the resolution
limit is the gradient of scalar function in the way the hydrodynamic
solver understands a gradient; gravitational and pressure resolutions 
for the gas are
again {\it exactly\/} equal as with the original SLH code. However,
since the dark matter particles are acted upon by the force $\hat f_i$,
the gravitational resolution for the dark matter component need not be
(but, of course, may be set) equal to the gravitational resolution 
for the gas component.

Now we turn to the details of implementing SLH-P$^3$M.
There exist two distinct ways to combine SLH hydro and P$^3$M gravity codes
(and many intermediate variants). 

The first one, which we call a
``maximally coupled'' code, implies that dark matter ``lives'' in 
quasi-Lagrangian space like in the original SLH code; then the P$^3$M
part is used to calculate forces acting on vertices of the mesh, where the
mass in a vertex is found by linear interpolation from the cell
centers and the mass of a cell is determined by the total mass 
of dark matter, gas,
and, possibly, galaxies in the cell. This approach requires calculating
the P$^3$M force only on $N_B$ mesh vertices, however, since the mesh is
tied to baryons, the maximally coupled version fails to follow accurately
places where the dark matter density differs significantly from the
gas density, like cores of big clusters.

The second variant, which we call a ``minimally coupled'' code,
allows dark matter to evolve in real space similar to existing SPH approaches.
The mesh is then considered to represent the baryonic gas only, and
only the gas mass is included in computing the cell mass. This variant
of the SLH-P$^3$M code does not suffer from the limitations of the minimally
coupled mode, but requires substantially more CPU time since computing of the
 gravitational force acting on both $N_B$ grid cells and $N_X$ dark matter 
particles is required. For the most common case $N_X=N_B$ this would require
twice the PM time and four times the PP time of the maximally coupled code.
When combined with the hydrodynamic part, this averages out to a minimally
coupled mode being approximately 2.5 times slower than the maximally coupled
mode. 

\placefig{
\begin{figure}
\insertfigure{\figdir/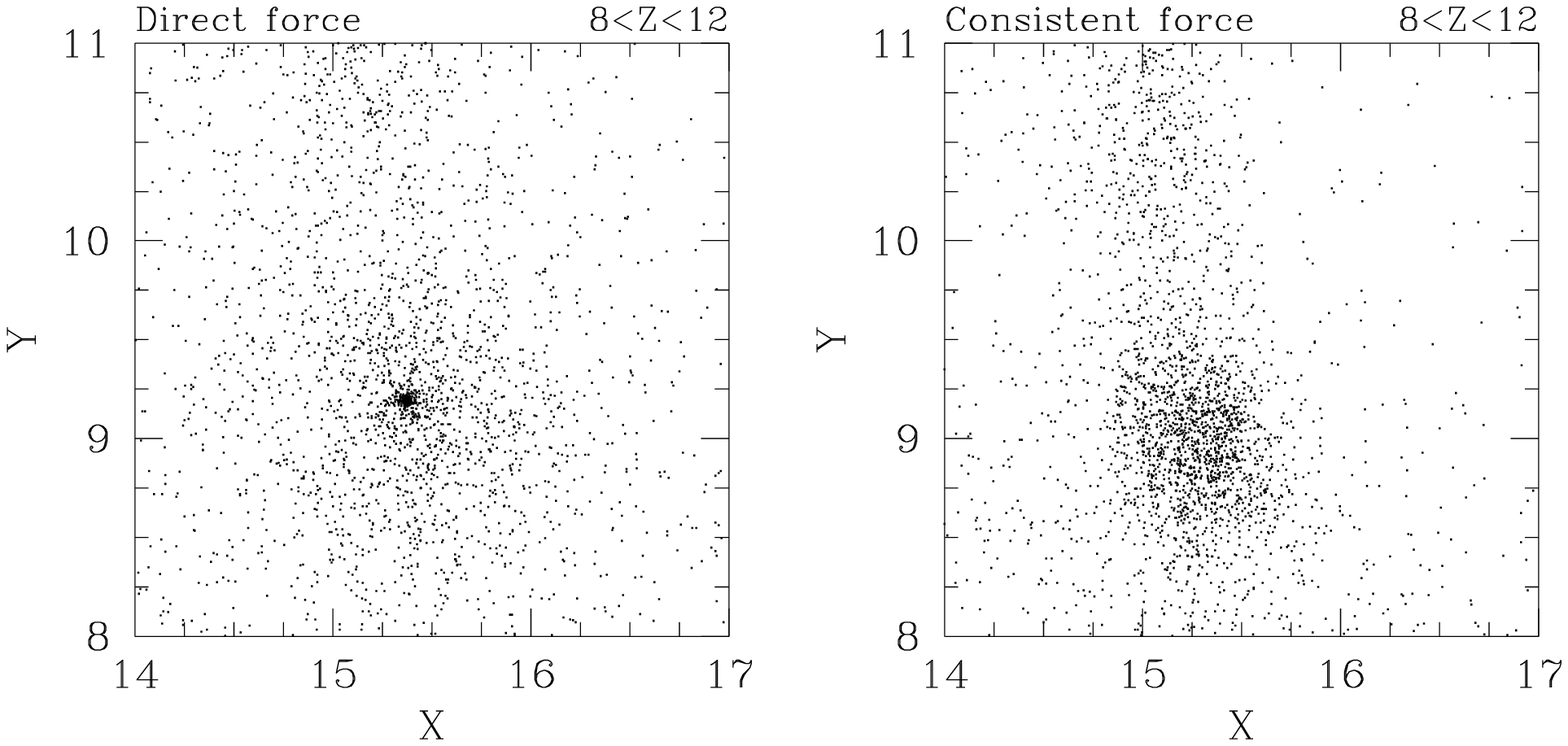}
\caption{\label{figPT}\capPT}
\end{figure}
}

We now proceed to testing the SLH-P$^3$M code. Since, in addition to the
mesh softening of SLH, the P$^3$M gravity solver introduces a Plummer
softening, the new SLH-P$^3$M code has two softening lengths. In what follows
we set these two softenings equal $1/10$ so that our simulation resolution will
not exceed that of a pure $N$-body template run; in principle, the 
gravitational force
acting on the gas can be computed with vanishing Plummer softening since
the mesh softening will prevent exceedingly small scales from being resolved.
Therefore, in the maximally coupled mode, when the dark matter, being
placed on the baryonic mesh, is subject to the mesh softening, the Plummer
softening can be assumed arbitrary small thus squeezing the last drop of
resolution from the simulation. One must note, however, that in that case
the resolution becomes even more anisotropic and nonuniform across the
simulation volume.

\placefig{
\begin{figure}
\insertfigure{\figdir/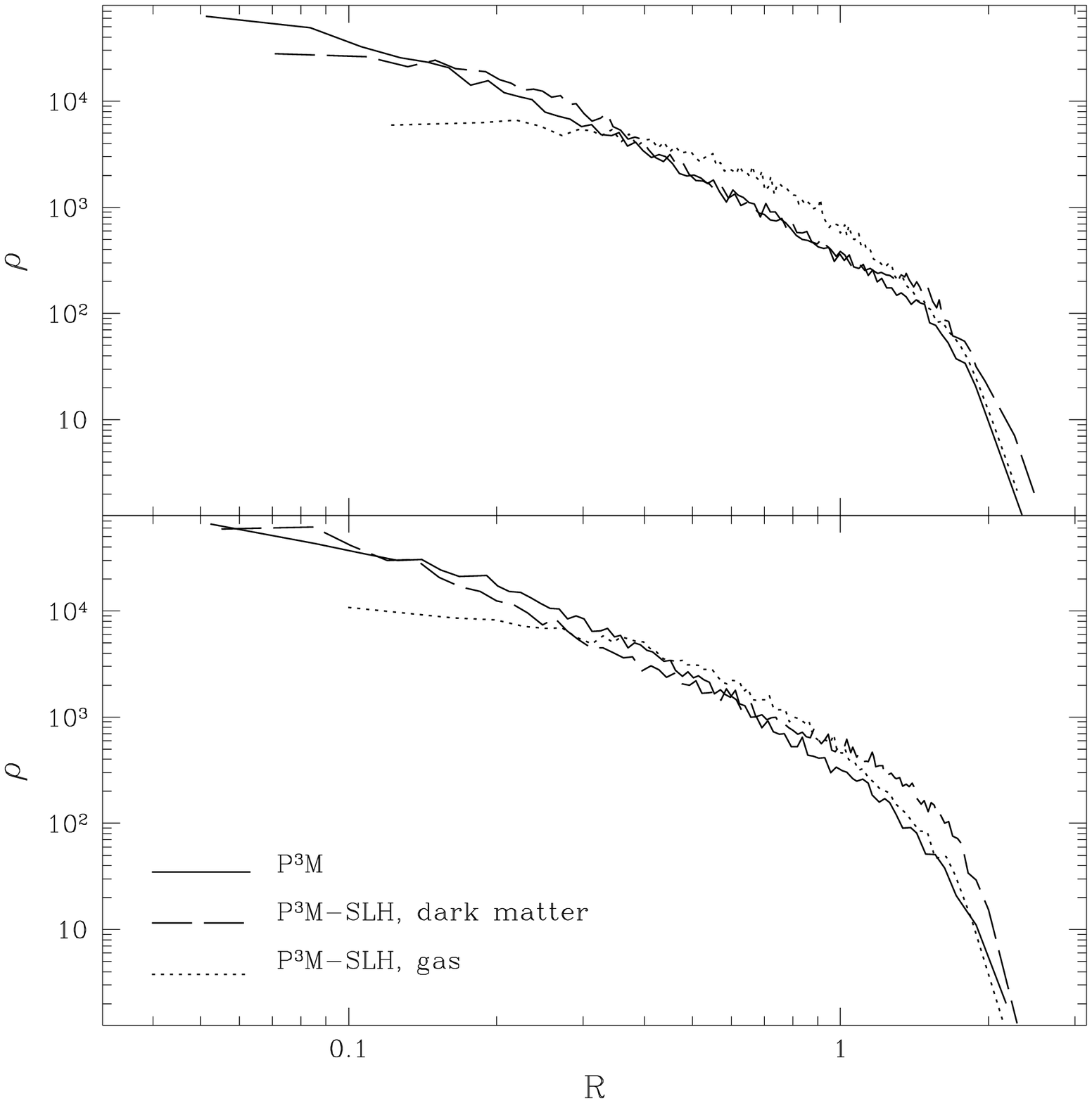}
\caption{\label{figPQ}\capPQ}
\end{figure}
}

First, we would like to demonstrate the importance of the
Consistency Condition. Figure \ref{figPT} shows the gas distribution of the
biggest clusters in two 32$^3$ 
baryons-only ($\Omega_X=0$) simulations
(cf.\ Fig.\ref{figSF}): the left panel shows the results for
the simulation where no Consistency Condition was applied, and the 
gravitational force acting on gas was directly provided by the P$^3$M
part of the code; the right panel shows the same cluster in the simulation
where the Consistency Condition of the form (\ref{ficonsistmax}) 
was incorporated
in the code (we remind the reader that the corresponding $N$-body result is
plotted in the lower right panel of Fig.\ref{figSF}). It is easily seen that
without the consistency condition most of the gas sinks to the very center
of the cluster thus violating energy conservation.
This effect of
gas sinking toward the cluster center can be solved without Consistency
Condition by simply smoothing the gravitational force on a scale of a
cell size; in that case the gravity force would have approximately the same
softening as the pressure force. This, nevertheless, would {\it not\/} solve
the problem since the cluster profile would still be significantly different
from the $N$-body results due to local nonconservation of energy. 

Since the original SLH method failed substantially only at the resolution
of a 64$^3$ mesh,
we ran a 64$^3$ $\Omega_X=0$ minimally-coupled SLH-P$^3$M simulation
(the case $\Omega_B=0$ coincides with the pure $N$-body result).
The profiles for the
two largest clusters in the simulation shown in Fig.\ref{figPQ} together
with the pure $N$-body result (the solid line). The dashed line shows the
dark matter profile and the dotted line shows the gas profile. While the
dark matter is fairly consistent with the pure $N$-body profiles, the
gas distribution has much larger cores. However, we can not attribute this
purely to lack of resolution: recent high resolution simulations have
also found this effect 
(Navarro, Frenk \& White\markcite{NFW95} 1995)\footnote{In 
real clusters gas cores are also 
much larger
than the dark matter cores as shown by comparison of X-ray pictures and
lensing reconstruction 
(cf.\ Miralda-Escude\markcite{M95} 1995); we do not use this fact as
an argument since it is not proved that our tests possess adequate 
resolution and physics to simulate real clusters.}. 
We thus conclude that the
agreement is acceptable and proceed to further test the new
SLH-P$^3$M code.

\section{Comparison with Other Codes}

\placefig{
\begin{figure}
\insertfigure{\figdir/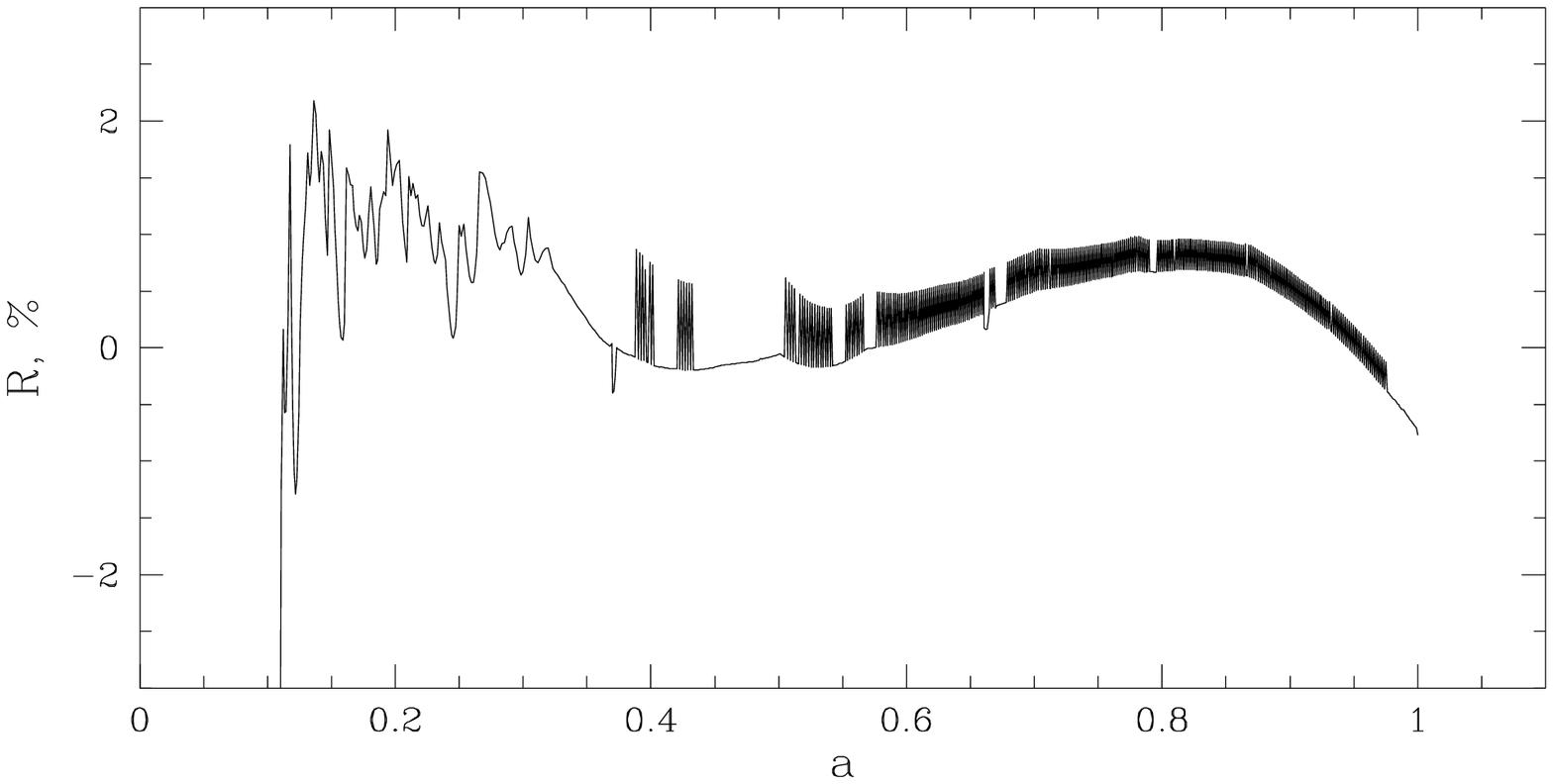}
\caption{\label{figEE}\capEE}
\end{figure}
}

\placefig{
\begin{figure}
\insertfigure{\figdir/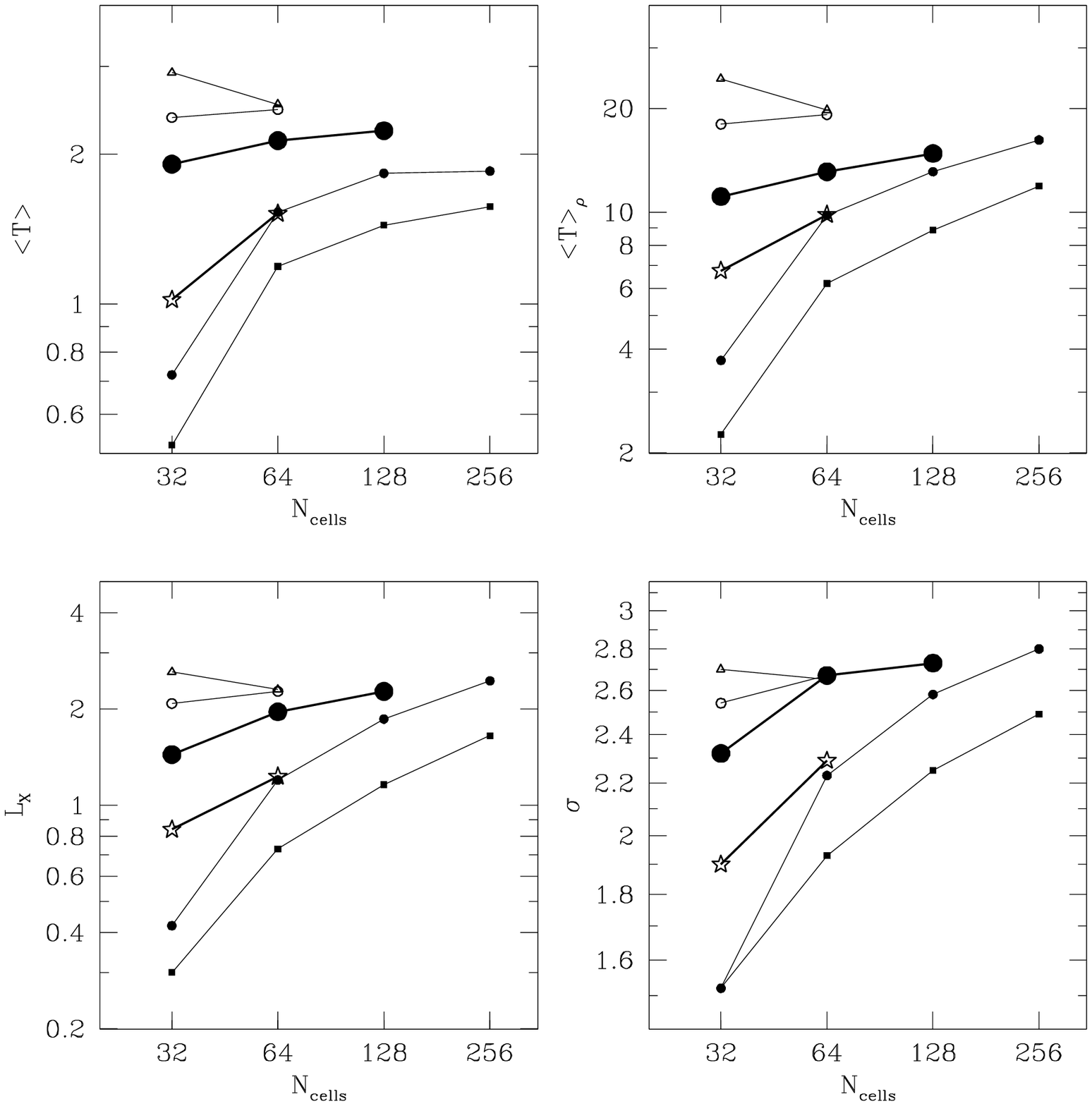}
\caption{\label{figCC}\capCC}
\end{figure}
}

In this section we 
repeat all the major tests of Kang et 
al.\markcite{Kea94} (1994) to assess differences between the repaired
SLH code and other hydrodynamic codes.
We performed three simulation with 32$^3$, 64$^3$, and 128$^3$ mesh
sizes with initial conditions of Kang et al.\markcite{Kea94} (1994)
\footnote{These are: Standard CDM model, $\Omega_b=1$, 
$64h^{-1}{\rm\,Mpc}$
box size; initial conditions were defined on a 64$^3$ mesh; for the 32$^3$
simulation every other value along each dimension was taken, and for 
128$^3$ simulation linear
interpolation was used to determine initial values on a 128$^3$ mesh.
Note, that the latter procedure in fact adds some small scale power to
the initial conditions and it is incorrect to say that the 128$^3$ 
SLH-P$^3$M simulation
and the 128$^3$ and 256$^3$ Eulerian simulations of 
Kang et al.\markcite{Kea94} (1994)
have the same initial conditions as the 64$^3$ simulations. In 
particular, the lack of convergence in the Kang et al.\markcite{Kea94} (1994)
Eulerian simulations is, due at least partially, to this effect.
However, we show results of a 128$^3$ simulation as a comparison with
larger Eulerian simulations since the initial condition for our 128$^3$
simulation are similar to 128$^3$ Eulerian simulations of
Kang et al.\markcite{Kea94} (1994).}. The total energy error in terms of
the quantity $R$ defined in Ryu et al.\markcite{Rea93} (1993) for the
SLH-P$^3$M\,64 run is plotted in Fig.\ref{figEE}. One can see that the 
total energy error fluctuates around 1\%, which is better than 
most existing cosmological hydrodynamic codes 
(see Table 1 of Kang et al.\markcite{Kea94} 1994 for comparison).

\placefig{
\begin{figure}
\fbox{\parbox{6in}{\begin{center}
\large This figure is too large for ASTRO-PH. 
The full paper
including color gif versions of figures 10-12 can be found at \\
ftp://arcturus.mit.edu/Preprints/slhp3m.ftp.tar.gz;\\ the file \\
slhp3m.complement.tar.gz \\ in the same directory contains
fig10-12 in b/w postscript and color gif versions only.
\end{center}}}
\caption{\label{figGS}\capGS}
\end{figure}
}

The final data for each simulation were rebinned onto a 
uniform 16$^3$ mesh, and integral
properties of the resulting 16$^3$ solution were calculated. 
Figure \ref{figCC} compares these against each other and 
against other codes shown in earlier work (Kang et al.\markcite{Kea94} 1994; 
Gnedin\markcite{G95} 1995).
Plotted on four
panels are the volume averaged temperature $\langle T\rangle$, 
the mass averaged
temperature $\langle T\rangle_\rho\equiv\langle\rho T\rangle
/\langle\rho\rangle$, the rms density fluctuation
$\sigma\equiv\langle\rho^2\rangle/\langle\rho\rangle^2-1$, 
and the total X-ray luminosity
$L_x=\sum\rho^2T^{1/2}$ in some arbitrary units. Four other cosmological codes
are presented together with the SLH code described in this paper; two
Eulerian hydrodynamic codes: the
Cen-Ostriker-Jameson code (Cen et al.\markcite{Cea90} 1990; filled squares),
the TVD code (Ryu et al.\markcite{Rea93} 1993; filled circles),
and two SPH codes: Tree-SPH of Hernquist \& Katz\markcite{HK89} 
(1989; open triangles) 
and P$^3$M-SPH of Evrard\markcite{E88} (1988; open circles). 
The self-evident notation used to designate these
codes is fully explained in Kang et al.\markcite{Kea94} (1994). 
The bold solid line with open
stars tracks the results of the original SLH code (which we shall hereafter
call SLH-MMG as an abbreviation to SLH-Moving Mesh Gravity)
 and the bold solid line
with solid circles shows the results of the new SLH-P$^3$M code. One can 
easily see that the SLH-P$^3$M code provides a major improvement 
in resolution
over the original SLH-MMG code, with resolution similar to 
or exceeding that of
the SPH codes 
for a 64$^3$ grid of particles while the temperature is similar to Eulerian
codes. The SLH-P$^3$M code, therefore, manages to reach the high 
resolution of
SPH codes while introducing less numerical heating.
The SLH-P$^3$M code is also a factor of 2 to 3 faster than 
a SPH code for the same resolution.

\placefig{
\begin{figure}
\fbox{\parbox{6in}{\begin{center}
\large This figure is too large for ASTRO-PH. 
The full paper
including color gif versions of figures 10-12 can be found at \\
ftp://arcturus.mit.edu/Preprints/slhp3m.ftp.tar.gz;\\ the file \\
slhp3m.complement.tar.gz \\ in the same directory contains
fig10-12 in b/w postscript and color gif versions only.
\end{center}}}
\caption{\label{figGD}\capGD}
\end{figure}
}

Let us now compare original (unrebinned) data for 
different codes. Fig.\ref{figGS}a
shows a one-cell-thick slice of original data of the SLH-P$^3$M\,64 run 
rebinned onto a
256$^3$ mesh (to facilitate the comparison with the TVD\,256 run); the slice
has $256^2$ points and is $0.25h^{-1}\mbox{Mpc}$ thick. Fig.\ref{figGS}b
 shows the 
same slice for SLH-P$^3$M\,128. Also shown are subslices containing one of the
largest clusters found in the whole slice. 

\placefig{
\begin{figure}
\fbox{\parbox{6in}{\begin{center}
\large This figure is too large for ASTRO-PH. 
The full paper
including color gif versions of figures 10-12 can be found at \\
ftp://arcturus.mit.edu/Preprints/slhp3m.ftp.tar.gz;\\ the file \\
slhp3m.complement.tar.gz \\ in the same directory contains
fig10-12 in b/w postscript and color gif versions only.
\end{center}}}
\caption{\label{figGT}\capGT}
\end{figure}
}

These slices are to be compared with
corresponding plots from Gnedin\markcite{G95} (1995), however, in order
to ease comparison, we plot subslices shown on the right
side of Fig.\ref{figGS}a,b together with corresponding
subslices for TVD\,256, PSPH\,64, and SLH-MMG\,64 simulations separately in
Fig.\ref{figGD} and Fig.\ref{figGT}; both figures show PSPH\,64 and TVD\,256 
results on the bottom; Fig.\ref{figGD} shows comparison between the original
SLH-MMG\,64 simulation (upper left panel) and the SLH-P$^3$M simulation (upper 
right panel) and Fig.\ref{figGT} plots SLH-P$^3$M\,64 and SLH-P$^3$M\,128
simulations in the upper row to demonstrate improvement and convergence.

One first notes that there are no significant differences visible between
the original SLH-MMG\,64 and SLH-P$^3$M\,64 results; the SLH-P$^3$M\,64 density
distribution resembles that of TVD\,256 slightly better, but shocks seem
to be more blurry since the higher resolution of the SLH-P$^3$M code leads
to fewer cells being left in voids and in the vicinity of shocks. The 
comparison between SLH-P$^3$M\,64 and SLH-P$^3$M\,128 shows little
difference in the density distribution but substantial improvement in the
shock thickness. Also the SLH-P$^3$M\,128 result has somewhat higher 
temperatures and shock extent because the gas was heated to higher
entropy by the extra small-scale power.

It is quite instructive to compare the SLH-P$^3$M 
results to SPH and TVD slices.
One can notice here that, except in the highest density regions, the
real resolution of an SPH simulation is low; it is hardly 
possible to identify filaments at all, whereas they are easily visible
in the SLH and TVD plots. In order to understand this somewhat unexpected
conclusion one must recall that the hydrodynamic quantities in the
SPH method are computed by averaging over, say, 30 nearest neighbors,
while in the SLH approach they defined in every cell and do not require
any averaging to compute. Since in the vicinity of filaments the
SLH mesh closely follows the gas flow, and, therefore, the hydrodynamical
description is close to Lagrangian, the SPH description of the gas is
similar to the SLH description being smoothed over 30 neighboring cells,
which roughly corresponds to $1/5-1/8$ of the whole slice size depending 
on the density. More than that, only at overdensities above 30 does the SPH
method acquire higher resolution than the Eulerian hydrodynamic code
with the same number of cells, and the TVD\,256 run becomes inferior in
spatial resolution to
the SPH\,64 run only at overdensities in excess of $30\times64\approx2000$.

In general, the SLH-P$^3$M\,128 run compares better to the TVD\,256 run 
than both SLH-P$^3$M\,64 and PSPH\,64. In some places the 
correspondence between
the former two runs is quite good, but also there are places where 
the two codes
differ substantially. We therefore conclude that in detail, differences
between existing hydrodynamic codes are still substantial, and much work
is required in order to achieve the same level of computational accuracy
and consistency as is observed within collisionless $N$-body codes.

\section{Summary}

We have demonstrated that Moving Mesh gravity solvers suffer from two serious
limitations: the shot-noise in high density regions and incorrect Green
functions. We have proposed a way to solve the former problem and failed to
completely identify the source of the latter. 
Since our Moving Mesh gravity solver can not
be trusted now, we have proceeded further by building a combined 
SLH-P$^3$M code,
where the SLH hydrodynamic solver is combined with the P$^3$M method for
computing gravity.

In order for a self-gravitating hydrodynamic code to conserve energy, a
special ``Consistency Condition'' ought to be satisfied. We have derived
the Consistency Condition for the SLH-P$^3$M code and showed what errors
would appear should this condition be violated. Ignoring this condition can
lead to serious errors in dense regions.

Finally, we have compared the SLH-P$^3$M code with existing cosmological
hydrodynamic codes including the original SLH code. The SLH-P$^3$M approach
offers a substantial increase in resolution, making it comparable to the
SPH codes in the highest density regions, while maintaining reasonable 
accuracy in resolving shocks and
hydrodynamic features. In particular, we have repaired the biggest problem
of the original SLH method: failure to simulate clusters accurately, which
stem out of a faulty Moving Mesh gravity solver.

We, therefore, conclude that the new SLH-P$^3$M code is a good numerical
tool to investigate the formation of nonlinear structure in the universe
including both gas and dark matter.

\acknowledgements

We are thankful to J.\ P.\ Ostriker for many fruitful discussions;
parallelization of the SLH-P$^3$M code for a shared memory multiprocessor was
improved with suggestions from F.\ Summers.
Supercomputer time was provided by the National Center for Supercomputing
Applications. This work was supported by NSF grant AST-9318185
awarded to the Grand Challenge Cosmology Consortium.

\appendix

\section{Softening Tensor}

At a general point in the flow the softening tensor is a function of the
deformation tensor $A^i_k$
\footnote{Note, that the definition of the deformation tensor used here
differs from that in the original SLH code of Gnedin\markcite{G95} (1995);
both definitions are similar but the one we used here generally leads to
a somewhat smoother mesh and to a slightly smaller number of time-steps 
in a simulation;
the resolution provided by both definitions is the same.}; let 
\begin{equation}
	\bar g^{ij} \equiv A^i_k \delta^{kl} A^j_l
\end{equation}
be the conjugate metric of quasi-Lagrangian space and let 
$\lambda_g(\bar g^{ij})$ be its eigenvalues.
Since $\bar g^{ij}$ is a real symmetric matrix,
one can introduce a unitary matrix $C^i_a$ such that
\begin{equation}
        \bar g^{ij} = C^i_a\mbox{diag}\left[\lambda_g\right]^{ab} C^j_b,
        \label{diagonalg}
\end{equation}
where $\mbox{diag}\left[\lambda_g\right]^{ab}$ denotes a diagonal matrix 
with $\lambda_g$ entries.

We can now introduce the softening tensor $\sigma^{ij}$ built according to
the following formula:
\begin{equation}
        \sigma^{ij} = C^i_a\mbox{diag}\left[\sigma(\sqrt{\lambda_g})
        \right]^{ab} C^j_b,
        \label{sigmaij}
\end{equation}
where $\sigma(\lambda)$ is defined by the following equation:
\begin{equation}
        \sigma(\lambda) = {1\over1 + (\lambda/\lambda_*)^2},
        \label{sigma}
\end{equation}
where $\lambda_*$ is a softening parameter.
In two opposite limits, when all three $\lambda_g$ approach zero or infinity,
the softening tensor tends to zero or $\delta^{ij}$ correspondingly.


\end{document}